\documentclass[
    ,final            % use final for the camera ready runs
  ]
  {aipproc}

\layoutstyle{6x9}

\begin{document}

\title{Thermal model for RHIC, part I: particle ratios and spectra
\thanks{Talk presented at II International Workshop on Hadron Physics,
{\em Effective Theories of Low-Energy QCD}, 25-29 September 2002,
University of Coimbra, Portugal}}

\author{Wojciech Florkowski and Wojciech Broniowski}{
  address={The H. Niewodnicza\'nski Institute of Nuclear Physics \\
ul. Radzikowskiego 152, 31-342 Krak\'ow, Poland}
}

\begin{abstract}
  A simple thermal model with single freeze-out and flow is used to
  analyze the ratios of hadron yields and the hadron transverse-mass
  spectra measured in $\sqrt{s_{NN}}$ = 130 GeV Au+Au collisions at
  RHIC. An overall very good agreement between the model predictions
  and the data is achieved for all measured hadron species including
  hyperons.
\end{abstract}

\maketitle

%%%%%%%%%%%%%%%%%%%%%%%%%%%%%%%%%%%%%%%%%%%%
%% MAINMATTER
%%%%%%%%%%%%%%%%%%%%%%%%%%%%%%%%%%%%%%%%%%%%

\section{Introduction}

The main features of the soft hadron production at RHIC, such as the ratios
of hadron abundances, the transverse-mass spectra, the elliptic flow, or
the HBT radii may be efficiently understood in the framework of a statistical
model which combines the standard thermal analysis of the hadron
ratios with a suitably parameterized expansion of matter at freeze-out
\cite{wbwfPRL,wbwfPRCstrange,ZAKOPANE}.  In the present paper we
outline the main assumptions of the model and concentrate on the
discussion of the particle ratios \cite{wfwbmm,mm} and the hadron
transverse-mass spectra \cite{RHICvSPS,wbHIRSCHEGG,wfEPIPHANY,wfQM02}.
The subsequent paper in these Proceedings \cite{CoimbraB} contains the
analysis of the elliptic flow and the pion HBT correlation radii.

The main ingredients of our approach are as follows: i) the chemical
freeze-out and the thermal (kinetic) freeze-out occur simultaneously,
ii) all hadronic resonances are included in the calculation of both
the hadron yields and the spectra, and iii) the freeze-out
hypersurface and the flow at freeze-out are defined by the simple
expressions inspired by the Bjorken model \cite{bjorken}. Below we
discuss in more detail these three points.

\subsection{Freeze-out}

Our approach includes a complete set of hadronic resonances in both
the calculation of the hadron abundances and the calculation of the
hadron spectra. Then, it turns out that the distinction between the
two freeze-outs \cite{2freezeouts} is not necessary.  Our analysis
showed \cite{wfwbmm} that the decays of the resonances, which are
initially present in a heat bath at the temperature of 165 MeV,
effectively lower the inverse slope parameters of the 
spectra by about 30-40 MeV. This is just the
desired effect which explains the typical difference between 
$T_{\rm chem}$ and $T_{\rm kin}$ (i.e., between the temperature at the
chemical freeze-out and the thermal freeze-out, respectively).  As a
consequence, we have found, at least for the RHIC data, that no extra
elastic rescattering is required in order to describe simultaneously
both the ratios and the spectra. In other words, we assume one
universal freeze-out taking place at \cite{wbwfPRL}
\begin{equation}
T_{\rm chem} = T_{\rm kin} \equiv T.
\label{sf}
\end{equation}
Recently, experimental hints have been found in favor of our
assumption (\ref{sf}). A successful reconstruction of the $K^*(892)^0$
states by the STAR experiment \cite{star_Kstar}, together with the very good
agreement between the measured yield of $K^*(892)^0$ and the
prediction of the thermal model \cite{ZAKOPANE,pbm_rhic} suggests a
picture with the short expansion time between the two freeze-outs.
Such a picture is natural if the production of particles
(hadronization) occurs in such conditions that neither elastic or
inelastic processes are effective. An example here is provided by the sudden
hadronization model of Ref. \cite{sudhadmod}.

\subsection{Resonances}

For thermal systems the contribution from high-lying (heavy) states is
damped by the exponential factor, $\,\exp(-m_{\perp}/T)$. This fact,
at the first sight, suggests that most of the resonances present in a
hadron gas at a moderate temperature may be neglected. However,
although the high-lying states are suppressed, their number increases
according to the Hagedorn hypothesis
\cite{hagedorn,2hagtem,hagbled,to_le_ra}, such that their net effect turns
out to be important. Indeed, hadronic resonances have been included in
numerous applications of the statistical models used in the studies of
the ratios of hadron multiplicities
\cite{pbm_rhic,pbm_sps,gg_appb,yen_gor,becattini_others} and the
effects connected with their decays are essential for the successful
description of the data. We note that only a quarter of the observed
pions at RHIC comes from the ``primordial'' pions present at
freeze-out, and the remaining three quarters are produced from the
decays of resonances.
 
In our approach, the same number of the resonances is included in the
calculation of the hadron ratios and in the calculation of the hadron
spectra. In this way, our theoretical spectra have always the correct
relative normalization. Moreover, we have worked out semi-analytic
formulas for the treatment of the resonance decays \cite{wfwbmm}. This
allows us to sum up exactly many small contributions, especially,
those appearing in the sequential decays. All two- and three-body
decays are taken into account with the branching ratios taken from the
tables. In the case of the three-body decays, the matrix elements are
approximated by a constant, hence only the phase-space effect is
included.

\subsection{Expansion}

In order to calculate the spectra we need to specify the expansion of
the matter at freeze-out; clearly, the Doppler effect due to flow modifies the
spectra and must be properly included. Our choice of the freeze-out
hypersurface and the four-velocity at freeze-out has been made in the
spirit of Refs.
\cite{bjorken,baym,milyutin,siemens1,schnedermann,BL,rischke1,scheibl}
and is defined by the two conditions:

\begin{equation}
\tau = \sqrt{t^2-r^2_x-r^2_y-r^2_z} = {\rm const.} \,\, ,
\label{tau}
\end{equation}
and
\begin{equation}
u^{\mu } =\frac{x^{\mu }}{\tau }=\frac{t}{\tau }\left(
1,\frac{r_{x}}{t},\frac{r_{y}}{t},\frac{r_{z}}{t}\right).
\label{umu}
\end{equation}
The constant in Eq. (\ref{tau}) will be later denoted simply by $\tau$.
In order to make the transverse size, 
\begin{equation}
\rho=\sqrt{r_x^2+r_y^2},
\label{rho}
\end{equation}
finite, we impose the condition 
\begin{equation}
\rho < \rho_{\rm max}. 
\label{rhomax}
\end{equation}
We note that the four-velocity (\ref{umu}) defining the hydrodynamic
flow at freeze-out is proportional to the coordinate (Hubble-like
expansion).  This form of the flow and the fact the the coordinates
$t$ and $r_z$ are not limited and appear in the boost-invariant
combination in (\ref{tau}) mean that our model is boost-invariant.  In
practical calculations it is convenient to introduce the following
parameterization \cite{BL}:
\begin{eqnarray}
t &=&\tau \cosh \alpha _{\parallel }\cosh \alpha _{\perp },\quad r_{z}=\tau
\sinh \alpha _{\parallel }\cosh \alpha _{\perp },  \nonumber \\
r_{x} &=&\tau \sinh \alpha _{\perp }\cos \phi ,\quad r_{y}=\tau \sinh \alpha
_{\perp }\sin \phi ,  \label{par}
\end{eqnarray}
where $\alpha _{\parallel }$ is the rapidity of the fluid element,
$v_{z}=r_z/t=\tanh \alpha _{\parallel }$, and $\alpha _{\perp }$ describes
the transverse size, $\rho =\tau \sinh \alpha _{\perp }$.
The transverse velocity is 
$v_\rho=\tanh \alpha_\perp/\cosh \alpha_\parallel$. 
The element of the hypersurface is defined as
\begin{equation}
d\Sigma_\mu = \epsilon_{\mu \alpha \beta \gamma}
{\partial x^\alpha \over \partial \alpha_\parallel}
{\partial x^\beta \over \partial \alpha_\perp}
{\partial x^\gamma \over \partial \phi} \, d\alpha_\parallel 
d\alpha_\perp d\phi, 
\label{sigma}
\end{equation}
where $x^0=t$, $x^1=r_x$, $x^2=r_y$, $x^3=r_z$, and $\epsilon_{\mu
\alpha \beta \gamma}$ is the Levi-Civita tensor. A straightforward
calculation yields
\begin{equation}
d\Sigma^\mu(x) = u^\mu(x)\, \tau ^{3} \, {\rm sinh}(\alpha _{\perp})
{\rm cosh}(\alpha _{\perp}) \, d\alpha _{\perp}
d\alpha _{\parallel } d\phi.
\label{prop}
\end{equation}
Equation (\ref{prop}) shows that the four-vectors $d\Sigma^\mu$ and
$u^\mu$ are parallel. In this case the spectra may be
obtained from the expression analogous to the Cooper-Frye 
\cite{cooperfrye1,cooperfrye2} formula
\begin{equation}
\frac{dN}{d^{2}p_{\perp }dy} =
\int p^{\mu }d\Sigma _{\mu }\ f\left(p\cdot u\right) ,
\label{Ni}
\end{equation}
but with the distribution $f$ which has collected the products of
resonance decays (for details see \cite{ZAKOPANE}). With
parameterization (\ref{par}) we can rewrite Eq. (\ref{Ni}) in the form
\begin{eqnarray}
\frac{dN}{2 \pi \, m_{\perp} \,dm_{\perp }dy} 
&=&\ \tau ^{3}\int_{-\infty }^{+\infty
}d\alpha _{\parallel }\int_{0}^{\rho _{\max }/\tau }{\rm sinh}  \alpha _{\perp
}d\left( {\rm sinh}  \alpha _{\perp }\right) \int_{0}^{2\pi }
d\xi \, p\cdot u \, f\left( p\cdot u\right) ,\nonumber \\
\label{dNi}
\end{eqnarray}
where
\begin{equation}
p\cdot u=m_{\perp }{\rm cosh} \alpha _{\parallel } {\rm cosh}  \alpha
_{\perp }-p_{\perp }\cos \xi \, {\rm sinh}  \alpha _{\perp }. \label{pu}
\end{equation}
One can notice that the spectrum (\ref{dNi}) is, as expected from the
assumed boost invariance, independent of the rapidity $y$.

\section{Ratios of hadron abundances}

In the case of the boost-invariant systems, the ratios of hadron
multiplicities at midrapidity, $dN/dy|_{y=0}$, are simply related to the
ratios of the local densities, $n_i$, since
\begin{equation}
\left . \frac{dN_i/dy}{dN_j/dy} \right |_{y=0}
=\frac{N_i}{N_j}=\frac{n_i}{n_j}. 
\label{dN}
\end{equation}
The first part of this equality follows from the boost invariance,
whereas the second part is a consequence of the factorization of the
volume of the system (this point is discussed in detail in Ref.
\cite{CoimbraB}).  Since the firecylinder formed at RHIC is
approximately boost-invariant (at least within one unit of rapidity at
$y=0$, which is sufficient for our considerations which concentrate
only on the central region), the ratios at zero rapidity measured for
various particles may be used to fit the thermal parameters of the
model. This is an important observation indicating that the ratios are
not sensitive to the particular form of expansion. Consequently, the
parameters of our model can be fixed in two steps: with help of the
ratios we first fix the thermodynamic parameters, and later with help
of the spectra we fix the two expansion parameters, $\tau$ and
$\rho_{\rm max}$.

The density of the $i$th hadron species is calculated from the
ideal-gas expression
\begin{eqnarray}
n_{i}&=&g_{i} \int d^3p \, f^{(i)}(p), \nonumber \\
f^{(i)}(p)&=&\frac{1}{(2\pi)^3} \left ( {\exp \left[
\left( E_{i}(p)-\mu _{B}B_{i}-\mu _{S}S_{i}-\mu
_{I}I_{i}\right) /T\right] \pm 1} \right )^{-1},  \label{ni}
\end{eqnarray}
where $g_{i}$ is the spin degeneracy, $ B_{i}$, $S_{i}$, and $I_{i}$
denote the baryon number, strangeness, and the third component of
isospin, and $E_{i}(p)=\sqrt{p^{2}+m_{i}^{2}}$. The quantities $ \mu
_{B},\mu _{S}$ and $\mu _{I}$ are the chemical potentials which enforce
the corresponding conservation laws. We recall that Eq.
(\ref{ni}) is used to calculate the ``primordial'' densities of stable
hadrons as well as of all resonances at freeze-out, which decay later on. 

Initially, the temperature, $T$, and the baryon chemical potential,
$\mu_{B}$, were fitted with the $\chi^2$ method to the 9 preliminary
experimental ratios of the hadron yields (for the list of the ratios
used in this fit and for more details of our analysis see Refs.
\cite{wfwbmm,mm}). The $\mu_{S}$ and $\mu_{I}$ were determined with
the conditions that the initial strangeness of the system is zero, and
the ratio of the baryon number to the electric charge is the same as
in the colliding nuclei. This procedure gave us $T=165\pm 7$ MeV and
$\mu _{B} = 41 \pm 5$ MeV.  In Table I we present the result of the
fit with a wider set of the up-to-date available hadronic ratios. It
yields $T=168\pm 5$ MeV and $\mu _{B} = 41 \pm 4$ MeV. It is
interesting to observe that the newly released data are fully consistent
with the thermal picture and only a small change of the thermodynamic
parameters follows when the new set of the ratios is used.

A characteristic feature of our fit is that the optimal temperature is
consistent with the value for the deconfinement phase transition
obtained from the QCD lattice simulations: $T_{c}=154\pm 8$~MeV for
three massless flavors and $T_{c}=173\pm 8$~MeV for two massless
flavors \cite{Karsch}).  This type of the behavior has been also found
in other statistical calculations \cite{pbm_rhic,becattini} and is
interpreted as the argument for a scenario in which the hadronic
ratios are fixed just in the hadronization process.

%%%%%%%%%%%%%%%%%%%%%%%%%%%%%%%%%%%%%%%%%%%%
%% SAMPLE TABLE
%%
%% Shows the use of \tablehead and \tablenote
%% macros
%%%%%%%%%%%%%%%%%%%%%%%%%%%%%%%%%%%%%%%%%%%%

\begin{table}[t]
\begin{centering}
\begin{tabular}{lrc}
\hline
& Model & Experiment \\ \hline 
\\
\multicolumn{3}{l}{Fitted thermal parameters}\\ %\hline
\\
$T$ [MeV] & 168$\pm 5$ &  \\ %\hline
$\mu _{B}$ [MeV] & \ 41$\pm 4$ &  \\ %\hline
$\mu _{S}$ [MeV] & \ \ \ \ \ 10 &  \\ %\hline
$\mu _{I}$ [MeV] & \ \ \ \ \ -1 &  \\ %\hline
$\chi^2/n$  & \ \ \ \ \ 0.6 &  \\ %\hline
\\
\multicolumn{3}{l}{Theoretical ratios and the data }\\ %\hline
\\
$\pi ^{-}/\pi ^{+}$ & $1.02$ & 
\begin{tabular}{ll}
$1.00\pm 0.02$ \cite{phobos_ratios_130}, & $0.99\pm 0.02$\cite{brahms_ratios_130}
\end{tabular} \\ %\hline
%%%%%%%%%%%%%%%%%%%%%%%%%%%%%%%%%% 
$\overline{p}/\pi ^{-}$ & $0.09$ & $0.08\pm 0.01$ \cite{star_ratios_harris} \\ %\hline
%%%%%%%%%%%%%%%%%%%%%%%%%%%%%%%%%%
$K^{-}/K^{+}$ & $0.92$ & 
\begin{tabular}{ll}
$0.92\pm 0.03$ \cite{star_strange_ratios}, & $0.93\pm 0.07$ \cite{phenix_ratios_ohnishi} \\ 
$0.91\pm 0.09$ \cite{phobos_ratios_130}, & $0.92\pm 0.06$ \cite{brahms_ratios_130}
\end{tabular} \\ %\hline
%%%%%%%%%%%%%%%%%%%%%%%%%%%%%%%%%%%
$K^{-}/\pi ^{-}$ & $0.16$ & $0.15\pm 0.02$ \cite{star_ratios_caines} \\ %\hline 
%%%%%%%%%%%%%%%%%%%%%%%%%%%%%%%%%%%
$K_{0}^{\ast }/h^{-}$ & 0.047 &  $0.042 \pm 0.011$ \cite{star_Kstar}  \\ 
$\overline{K_{0}^{\ast }}/h^{-}$ & $0.042$ &  $0.039 \pm 0.011$ \cite{star_Kstar} \\ 
%%%%%%%%%%%%%%%%%%%%%%%%%%%%%%%%%%%
$\overline{p}/p$ & $0.65$ & 
\begin{tabular}{ll}
$0.61\pm 0.07$ \cite{star_ratios_harris}, & $0.64\pm 0.08$ \cite{phenix_ratios_ohnishi} \\ 
$0.60\pm 0.07$ \cite{phobos_ratios_130}, & $0.61\pm 0.06$ \cite{brahms_ratios_130}
\end{tabular} \\ %\hline
%%%%%%%%%%%%%%%%%%%%%%%%%%%%%%%%%%%
$\overline{\Lambda }/\Lambda $ & $0.69$ & $0.71\pm 0.04$ \cite{star_strange_ratios} \\ %\hline
%%%%%%%%%%%%%%%%%%%%%%%%%%%%%%%%%%%
$\overline{\Xi }/\Xi $ & $0.77$ & $0.83\pm 0.06$ \cite{star_strange_ratios} \\ %\hline
%%%%%%%%%%%%%%%%%%%%%%%%%%%%%%%%%% 
$\phi/h^-$ & $0.020$ & $0.021 \pm 0.001$ \cite{star_phi} \\ %\hline
%%%%%%%%%%%%%%%%%%%%%%%%%%%%%%%%%%
$\phi/K^-$ & $0.15$ & $0.13 \pm 0.03$ \cite{star_phi} \\ %\hline
%%%%%%%%%%%%%%%%%%%%%%%%%%%%%%%%%%
$\Lambda/p$ & 0.47 & $0.49 \pm 0.03$ \cite{star_Lambda,star_antiprotons} \\ %\hline
%%%%%%%%%%%%%%%%%%%%%%%%%%%%%%%%%%
$\Omega^-/h^-$ & 0.0011 & $0.0012 \pm 0.005$ \cite{star_Omega} \\ %\hline
%%%%%%%%%%%%%%%%%%%%%%%%%%%%%%%%%%
$\Xi^-/\pi^-$ & 0.0072 & $0.0088 \pm 0.0020$ \cite{star_Xi} \\ %\hline
%%%%%%%%%%%%%%%%%%%%%%%%%%%%%%%%%%%
$\Omega^+/\Omega^-$ & 0.86 & $0.95 \pm 0.16$ \cite{star_strange_ratios} \\ %\hline
\end{tabular}
\caption{Optimal thermal parameters and the ratios of hadron multiplicities
at zero rapidity.}
\end{centering}
\label{tab:fit}
\end{table}

\begin{figure}[t]
\includegraphics[height=.8\textheight]{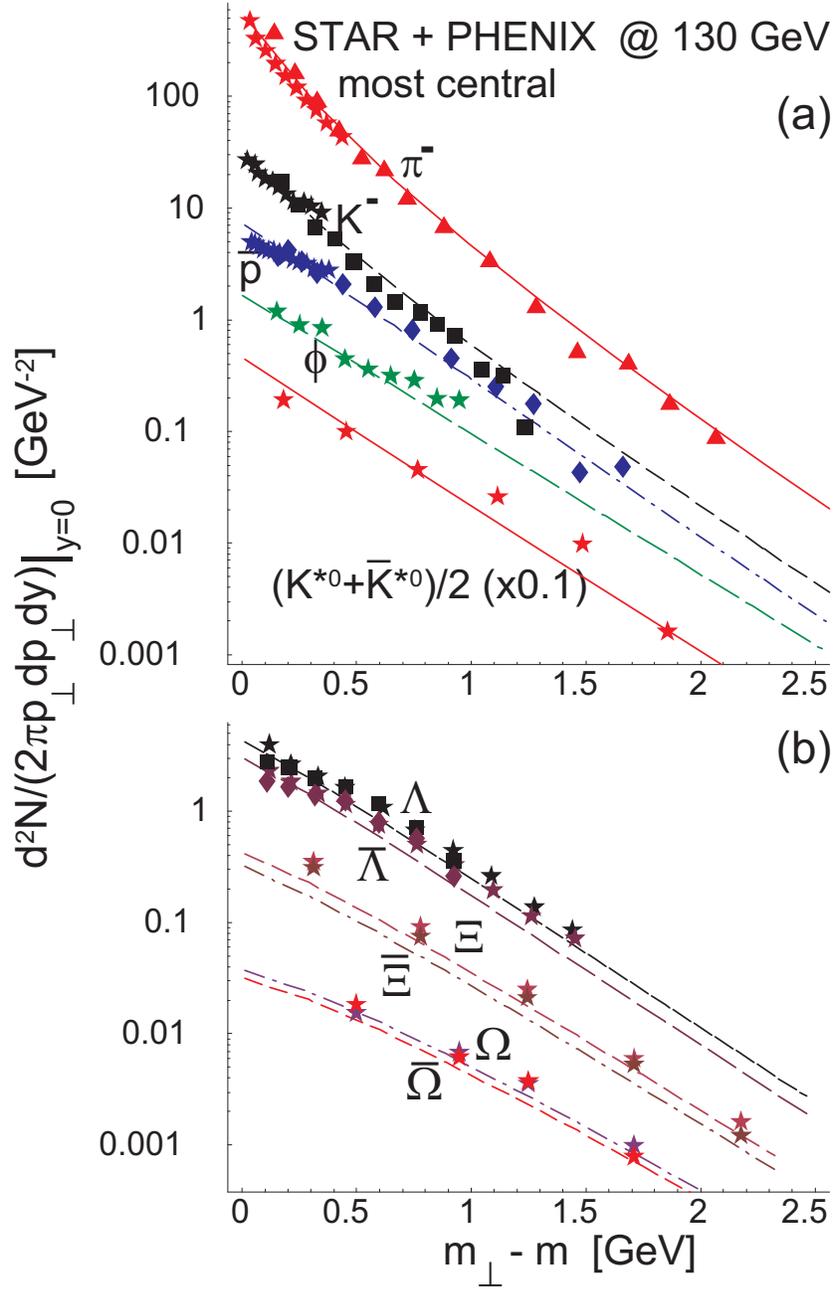}
\caption{The transverse-mass spectra at midrapidity. The data from
STAR are denoted by asterisks, other symbols are used to denote the PHENIX data.
All spectra are for most central collisions}
\end{figure}

\section{Transverse-mass spectra}

Having fixed the two independent thermodynamic parameters of the
model, we can use Eq. (\ref{dNi}) to fit the transverse-mass
spectra and to fix the two remaining geometric parameters of the
model. This method has been initially applied in the study of the
spectra of pions, kaons and protons, which gave us, for the most
central events, the following values of the size parameters:
\begin{equation}
\tau = 7.66 \hbox{ fm}, \hspace{1.5cm} 
\rho_{\rm max} = 6.69 \hbox{ fm}.
\label{param}
\end{equation}
In Fig. 1 we show our results for all up-to-now available spectra at
$\sqrt{s_{NN}}=130$~GeV for the most-central collisions. In the upper
part of Fig. 1 we show the spectra of pions, kaons, and antiprotons
(used earlier to determine the geometric parameters) and the predicted
spectra of the $\phi$ and $K^*(892)^0$ mesons.  The predicted spectrum
of the $\phi$ mesons agrees very well with the measurement
\cite{star_phi}, with model curve crossing five out of the nine data
points.  The $\phi$ meson production is of a particular interest in
relativistic heavy-ion collisions, since its spectrum reflects the
initial temperature of the hadronic system. This is because its
interaction with the hadronic environment is negligible. Moreover, it
does not receive any contribution from resonance decays. Thus, the
agreement of the model and the data supports the idea of one universal
freeze-out.  The upper part of Fig. 1 also shows the averaged spectrum
of $K^*$ resonances, with the data from Ref.  \cite{star_Kstar}.  Once
again we observe a very good agreement between the model curve and the
experimental points. As already mentioned above, the successful
description of both the yield and the spectrum of $K^*(892)^0$ mesons
supports the concept of the thermal description of hadron production
at RHIC, and brings evidence for a very small interval between chemical and
thermal freeze-outs.

In the lower part of Fig. 1  we show the predictions of the
model for the spectra of hyperons. Again, in view of the fact that no
extra parameters are introduced here and no refitting has been
performed, the agreement is impressive. We note that the preliminary
\cite{star_Xi_old} data for the $\Xi$'s used in the figure were
subsequently updated, which resulted in the reduction of the data by
about a factor of 2. This correction makes our agreement with the data
even better. The data accumulated at lower energies at SPS showed that
the slope of the $\Omega$ hyperon was much steeper than for other
particles \cite{cern_Omega}. On the contrary, in the case of RHIC the
model predictions for the $\Omega$ are as good as for the other
hadrons.  Since the $\Omega$ contains three strange quarks, it is most
sensitive for modifications of the simple thermal model used here,
{\em e.g.} the use of canonical instead of the grand-canonical
ensemble. The agreement of Fig. 1  does not support the need for
inclusion of these effects.

\section{Summary}

The success of our model in reproducing the hadron ratios and the
transverse-mass spectra indicates that the particles are indeed
produced thermally. The model has only four parameters and describes
the data with surprising accuracy.  Since our approach uses hadronic
degrees of freedom and starts at freeze-out, important theoretical
questions concerning the earlier stages of the evolution cannot be
addressed in this framework. We think, however, that our results
constrain any more microscopic descriptions of the evolution of the
matter formed in ultra-relativistic heavy-ion collisions. 
Further applications of the model aiming at the description of the 
elliptic flow and the HBT pion radii will be presented in the next
contribution to these Proceedings \cite{CoimbraB}.

%%%%%%%%%%%%%%%%%%%%%%%%%%%%%%%%%%%%%%%%%%%%%%%%
%% BACKMATTER
%%%%%%%%%%%%%%%%%%%%%%%%%%%%%%%%%%%%%%%%%%%%%%%%

\begin{theacknowledgments}
 Supported in part by the Polish State Committee for
Scientific Research, grant 2 P03B 09419. WB acknowledges the
support of PRAXIS XXI/BCC/429/94.
\end{theacknowledgments}

\hyphenation{Post-Script Sprin-ger}


\begin{thebibliography}{47}
\expandafter\ifx\csname natexlab\endcsname\relax\def\natexlab#1{#1}\fi
\providecommand{\enquote}[1]{``#1''}
\expandafter\ifx\csname url\endcsname\relax
  \def\url#1{\texttt{#1}}\fi
\expandafter\ifx\csname urlprefix\endcsname\relax\def\urlprefix{URL }\fi

\bibitem[Broniowski and Florkowski(2001)]{wbwfPRL}
Broniowski, W., and Florkowski, W., \emph{Phys. Rev. Lett.}, \textbf{87},
  272302 (2001).

\bibitem[Broniowski and Florkowski(2002{\natexlab{a}})]{wbwfPRCstrange}
Broniowski, W., and Florkowski, W., \emph{Phys. Rev. C}, \textbf{65}, 064905
  (2002{\natexlab{a}}).

\bibitem[Broniowski et~al.(hep-ph/0209286)]{ZAKOPANE}
Broniowski, W., Baran, A., and Florkowski, W., 
\emph{Acta Phys. Pol. B}, \textbf{33}, 4235 (2002).

\bibitem[Florkowski et~al.(2002)]{wfwbmm}
Florkowski, W., Broniowski, W., and Michalec, M., \emph{Acta Phys. Pol. B},
  \textbf{33}, 761 (2002).

\bibitem[Michalec(2001)]{mm}
Michalec, M., \emph{{Thermal description of particle production in
  ultra-relativistic heavy-ion collisions}}, Ph.D. thesis, Institute of Nuclear
  Physics, ul. Radzikowskiego 152, 31-342 Krak\'ow, Poland (2001), available as
  nucl-th/0112044.

\bibitem[Broniowski and Florkowski(2002{\natexlab{b}})]{RHICvSPS}
Broniowski, W., and Florkowski, W., \emph{Acta Phys. Pol. B}, \textbf{33}, 1935
  (2002{\natexlab{b}}).

\bibitem[Broniowski and Florkowski(2002{\natexlab{c}})]{wbHIRSCHEGG}
Broniowski, W., and Florkowski, W., \enquote{Thermal model at RHIC: particle
  ratios and $p_\perp$ spectra,} in \emph{Ultrarelativistic Heavy-Ion
  Collisions}, edited by M.~Buballa, W.~N\"orenberg, B.-J. Schaefer, and
  J.~Wambach (GSI, Darmstadt) Hirschegg, Austria, 2002{\natexlab{c}}, p. 146,
hep-ph/0202059.

\bibitem[Florkowski and Broniowski(2002)]{wfEPIPHANY}
Florkowski, W., and Broniowski, W., \emph{Acta Phys. Pol. B}, \textbf{33}, 1629
  (2002).

\bibitem[Florkowski and Broniowski(nucl-th/0208061)]{wfQM02}
Florkowski, W., and Broniowski, W., \enquote{Thermal description of
  transverse-momentum spectra at RHIC,} in \emph{Proceedings of Quark Matter
  2002 Conference}, Nucl. Phys. A in print, nucl-th/0208061.

\bibitem[Broniowski et~al.(2002)]{CoimbraB}
Broniowski, W., Baran, A., and Florkowski, W., \enquote{Thermal model at RHIC,
  part II: elliptic flow and HBT radii,} following paper.

\bibitem[Bjorken(1983)]{bjorken}
Bjorken, J.~D., \emph{Phys. Rev. D}, \textbf{27}, 140 (1983).

\bibitem[Heinz(1999)]{2freezeouts}
Heinz, U., \emph{Nucl. Phys. A}, \textbf{661}, 140c (1999).

\bibitem[Xu(2002)]{star_Kstar}
Xu, Z., \emph{{\rm STAR Collaboration}, Nucl. Phys. A}, \textbf{698}, 607c
  (2002).

\bibitem[Braun-Munzinger et~al.(2001)]{pbm_rhic}
Braun-Munzinger, P., Magestro, D., Redlich, K., and Stachel, J., \emph{Phys.
  Lett. B}, \textbf{518}, 41 (2001).

\bibitem[Rafelski and Letessier(2000)]{sudhadmod}
Rafelski, J., and Letessier, J., \emph{Phys. Rev. Lett.}, \textbf{85}, 4695
  (2000).

\bibitem[Hagedorn(1965)]{hagedorn}
Hagedorn, R., \emph{Suppl. Nuovo Cim.}, \textbf{3}, 147 (1965).

\bibitem[Broniowski and Florkowski(2000)]{2hagtem}
Broniowski, W., and Florkowski, W., \emph{Phys. Lett. B}, \textbf{490}, 223
  (2000).

\bibitem[Broniowski(2002)]{hagbled}
Broniowski, W., \enquote{Distinct Hagedorn temperatures from the particle
  spectra: a higher one for mesons, a lower one for baryons,} in
  \emph{Few-Quark Problems}, edited by B.~Golli, M.~Rosina, and S.~{\v S}irca,
  Bled, Slovenia, 2002, p.~14, hep-ph/0008112.

\bibitem[Tounsi et~al.(1994)]{to_le_ra}
Tounsi, A., Letessier, J., and Rafelski, J., \enquote{Hadronic matter equation
  of state and the hadron mass spectrum,} in \emph{Hot Hadronic Matter},
  Divonne-les-Bains, France, 1994, p. 105.

\bibitem[Braun-Munzinger et~al.(1999)]{pbm_sps}
Braun-Munzinger, P., Heppe, I., and Stachel, J., \emph{Phys. Lett. B},
  \textbf{465}, 15 (1999).

\bibitem[Ga\'zdzicki and Gorenstein(1999)]{gg_appb}
Ga\'zdzicki, M., and Gorenstein, M.~I., \emph{Acta Phys. Pol. B}, \textbf{30},
  2705 (1999).

\bibitem[Yen and Gorenstein(1999)]{yen_gor}
Yen, G.~D., and Gorenstein, M.~I., \emph{Phys. Rev. C}, \textbf{59}, 2788
  (1999).

\bibitem[Becattini et~al.(2001)]{becattini_others}
Becattini, F., Cleymans, J., Keranen, A., Suhonen, E., and Redlich, K.,
  \emph{Phys. Rev. C}, \textbf{64}, 024901 (2001).

\bibitem[Baym et~al.(1983)]{baym}
Baym, G., Friman, B., Blaizot, J.~P., Soyeur, M., and Czy\.z, W., \emph{Nucl.
  Phys. A}, \textbf{407}, 541 (1983).

\bibitem[Milyutin and Nikolaev(1998)]{milyutin}
Milyutin, P., and Nikolaev, N.~N., \emph{Heavy Ion Phys.}, \textbf{8}, 333
  (1998).

\bibitem[Siemens and Rasmussen(1979)]{siemens1}
Siemens, P.~J., and Rasmussen, J., \emph{Phys. Rev. Lett.}, \textbf{42}, 880
  (1979).

\bibitem[Schnedermann et~al.(1993)]{schnedermann}
Schnedermann, E., Sollfrank, J., and Heinz, U., \emph{Phys. Rev. C},
  \textbf{48}, 2462 (1993).

\bibitem[Cs\"{o}rg\H{o} and L\"orstad(1996)]{BL}
Cs\"{o}rg\H{o}, T., and L\"orstad, B., \emph{Phys. Rev. C}, \textbf{54}, 1390
  (1996).

\bibitem[Rischke and Gyulassy(1996)]{rischke1}
Rischke, D.~H., and Gyulassy, M., \emph{Nucl. Phys. A}, \textbf{597}, 701
  (1996).

\bibitem[Scheibl and Heinz(1999)]{scheibl}
Scheibl, R., and Heinz, U., \emph{Phys. Rev. C}, \textbf{59}, 1585 (1999).

\bibitem[Cooper and Frye(1974)]{cooperfrye1}
Cooper, F., and Frye, G., \emph{Phys. Rev. D}, \textbf{10}, 186 (1974).

\bibitem[Cooper et~al.(1975)]{cooperfrye2}
Cooper, F., Frye, G., and Schonberg, E., \emph{Phys. Rev. D}, \textbf{11}, 192
  (1975).

\bibitem[Karsch(2002)]{Karsch}
Karsch, F., \emph{Nucl. Phys. A}, \textbf{698}, 199c (2002).

\bibitem[Becattini(2002)]{becattini}
Becattini, F., \emph{J. Phys. G}, \textbf{28}, 1553 (2002).

\bibitem[Back(2001)]{phobos_ratios_130}
Back, B.~B., et al., \emph{{\rm PHOBOS Collaboration}, Phys. Rev. Lett.}, \textbf{87},
  102301 (2001).

\bibitem[Bearden(2002)]{brahms_ratios_130}
Bearden, I.~G., \emph{{\rm BRAHMS Collaboration}, Nucl. Phys. A}, \textbf{698},
  667c (2002).

\bibitem[Harris(2002)]{star_ratios_harris}
Harris, J., \emph{{\rm STAR Collaboration}, Nucl. Phys. A}, \textbf{698}, 64c
  (2002).

\bibitem[sta(2002)]{star_strange_ratios} Adams, J., et al., 
\emph{{\rm STAR Collaboration, nucl-ex/0211024}}.

\bibitem[Ohnishi(2002)]{phenix_ratios_ohnishi}
Ohnishi, H., \emph{{\rm PHENIX Collaboration}, Nucl. Phys. A}, \textbf{698},
  659c (2002).

\bibitem[Caines(2002)]{star_ratios_caines}
Caines, H., \emph{{\rm STAR Collaboration}, Nucl. Phys. A}, \textbf{698}, 112c
  (2002).

\bibitem[Adler(2002{\natexlab{a}})]{star_phi}
Adler, C., et al., \emph{{\rm STAR Collaboration}, Phys. Rev. C}, \textbf{65}, 041901
  (2002{\natexlab{a}}).

\bibitem[Adler(2002{\natexlab{b}})]{star_Lambda}
Adler, C., et al., \emph{{\rm STAR Collaboration}, Phys. Rev. Lett.}, \textbf{89},
  092301 (2002{\natexlab{b}}).

\bibitem[Adler(2001)]{star_antiprotons}
Adler, C., et al., \emph{{\rm STAR Collaboration}, Phys. Rev. Lett.}, \textbf{87},
  262302 (2001).

\bibitem[Suire(2002)]{star_Omega}
Suire, C., \emph{{\rm STAR Collaboration, nucl-ex/0211017}}.

\bibitem[Castillo(2002)]{star_Xi}
Castillo, J., \emph{{\rm STAR Collaboration, nucl-ex/0210032}}.

\bibitem[Castillo(2002)]{star_Xi_old}
Castillo, J., \emph{{\rm STAR Collaboration}, J. Phys. G.}, \textbf{28}, 1987
  (2002).

\bibitem[Antinori(2001)]{cern_Omega}
Antinori, C., et al., \emph{{\rm WA97 Collaboration}, J. Phys. G.}, \textbf{27}, 375
  (2001).

\end{thebibliography}
\end{document}